\documentclass[prl,twocolumn,superscriptaddress,floatfix,showpacs]{revtex4-1}
\usepackage{graphics,amssymb,amsmath,epsfig,color}

\newcommand{\be}{\begin{equation}} \newcommand{\ee}{\end{equation}}
\newcommand{\bea}{\begin{eqnarray}} \newcommand{\eea}{\end{eqnarray}}

\begin{document}

\title{Are Percolation Transitions always Sharpened by Making Networks Interdependent?}
\author{Seung-Woo Son} \affiliation{Complexity Science Group, University of Calgary, Calgary T2N 1N4, Canada}
\author{Peter Grassberger} \affiliation{Complexity Science Group, University of Calgary, Calgary T2N 1N4, Canada}
\author{Maya Paczuski} \affiliation{Complexity Science Group, University of Calgary, Calgary T2N 1N4, Canada}
\date{\today}

\begin{abstract}
We study a model for coupled networks introduced recently by Buldyrev {\it et al.}, 
Nature {\bf 464}, 1025 (2010), where each node has to be connected to others via 
two types of links to be viable. Removing a critical fraction of nodes leads to a
percolation transition that has been claimed to be more abrupt than that for uncoupled
networks. Indeed, it was found to be discontinuous in all cases studied. Using an 
efficient new algorithm we verify that the transition is discontinuous for coupled 
Erd\"os-R\'enyi networks, but find it to be continuous for fully interdependent diluted 
lattices. In 2 and 3 dimension, the order parameter exponent $\beta$ is {\it larger} 
than in ordinary percolation, showing that the transition is {\it less} 
sharp, i.e. further from discontinuity, than for isolated networks. Possible consequences 
for spatially embedded networks are discussed.

\end{abstract}


\maketitle

While the theoretical study of {\it single} networks has exploded during the last years, relatively little 
work has been devoted to the study of interdependent networks. This is in stark contrast to the 
abundance of coupled networks in nature and technology -- one might e.g. think of people connected 
by telephone calls, by roads, by their work relationships, etc. For single networks it is well known 
that removing nodes can lead to cascades where other nodes become dysfunctional too \cite{Motter}, 
and deleting a sufficient fraction of nodes leads to the disappearance of the giant connected cluster.
If the network is already close to the transition point, deleting a single node can lead to 
an infinite cascade similar to the outbreak of a large epidemic in a population.

Assume now that all nodes have to be connected via different types of links in order to remain functional.
It was argued in \cite{Buldy} that in such cases the cascades of failure triggered by 
removing single nodes should be greatly enhanced, and that the transition between existence and 
non-existence of a giant cluster of functional nodes should become discontinuous. 
This claim was backed by a mean field theory that becomes exact for locally 
tree-like networks (e.g. large sparse Erd\"os-R\'enyi (ER) networks), and by numerical 
simulations for various types of network topologies. 
In the present paper we show that this view
is not entirely correct: For fully interdependent diluted $d$-dimensional lattices, the transition is 
not only continuous, but it is {\it less sudden} than the ordinary percolation (OP) transition 
for isolated lattices and represents a new universality class. 

The problem is best illustrated by an actual case discussed in \cite{Buldy}, which concerns an electric 
power blackout in Italy in September 2003 \cite{Rosato}. According to \cite{Buldy} (see also
\cite{Parshani,Havlin}),
the event was possibly triggered by the failure of a single node $i_0$ in the electricity network. 
Nodes in a power networks are in general also linked by a
telecommunication network (TN) and need to receive information about the status of the 
other nodes. In the present case, presumably some nodes in the TN failed,
because they were not supplied with power. This then led to the failure of more power stations
because they did not receive the necessary information from $i_0$, of more TN nodes 
because they were not supplied with electric power, etc. The ensuing cascade finally affected 
the entire power grid.

The crucial point here is that each node has to be connected to two distinct networks that provide 
different services, in order to be viable. At the same time nodes act as bridges to bring 
supply to other nodes.  If a node gets disconnected from one network, it no longer can 
function and looses also its ability to serve as a connector in the other. 
The claim in \cite{Buldy}, to be scrutinized here, is that these cascades
of failure are much more abrupt in interdependent networks than in isolated ones, leading to 
much sharper transitions.

In a single network, the existence of an ``infinite" cluster of nodes, making possible the 
outbreak of a large epidemic, is described by OP. 
Whether such a large outbreak can happen
depends on the average connectivity of the network, characterized by some parameter $p$. If 
$p$ is below a critical value $p_c$, no infinite epidemic can occur, while it occurs with 
probability $P>0$ if $p>p_c$. 
For $p$ slightly above $p_c$, both $P$ and the relative size of the epidemic in
a large but finite population scale $ \sim (p-p_c)^\beta$, where the 
{\it order parameter exponent} $\beta$ depends on the 
topology of the network. For ER networks $\beta=1$, while for randomly diluted $d$-dimensional
lattices $\beta$ depends on $d$, with $\beta(d=2) = 5/36\approx 0.1389$ \cite{Stauffer} and
 $\beta(d=3) = 0.4170(3)$ \cite{Deng}. In all these cases $\beta>0$, meaning that the 
transition is continuous. A discontinuous transition, as found in \cite{Buldy,Havlin},
would correspond to $\beta=0$.

Discontinuous percolation transitions have recently been claimed to exist in several
other models \cite{Achliopt,Herrmann-Manna-Chen}, including 
{\it explosive percolation} \cite{Achliopt}. The numerical evidence
for discontinuity given in \cite{Achliopt} was supported in numerous papers. It became
clear only recently that the transition is actually continuous, although with small
$\beta$ and with unusual finite size behavior \cite{Costa-Grass-Riordan-Lee}. In view of the
difficulty to distinguish numerically between a truly discontinuous transition and a 
continuous one with very small $\beta$, we decided to perform more precise simulations.

The algorithm used in \cite{Buldy} follows in detail the 
cascades triggered by removing nodes and, as a result, does
not allow one to study large networks with high statistics. In our simulations,
instead of removing nodes, we add nodes one by one.
Using a modification of the fast Newman-Ziff algorithm \cite{Newman}, this gives a 
code which no longer follows entire cascades, as they are broken up into 
short sub-cascades, and gluing them together would make the algorithm slow again. But it 
allowed us to obtain high statistics for reasonably large systems. 

The model is formally defined as follows: Start with a single set $\cal N$ of $N$ nodes and with
two networks $\cal A$ and $\cal B$ that are obtained by linking these nodes (notice that 
$\cal A$ and $\cal B$ need not be connected, and indeed some nodes in $\cal N$
may be not connected at all, in which case $\cal A$ and $\cal B$ make use only of subsets
of $\cal N$; also we do not demand that all links in $\cal A$ and $\cal B$ are different).
Typically, we construct $\cal A$ and $\cal B$ by starting with a dense network and deleting randomly 
links from it, keeping links only with probability $q<1$. In this way, ER networks are constructed
by starting with a complete graph and keeping only $L = qN(N-1)/2$ links. Alternatively, diluted 
regular $d$-dimensional lattices are obtained by starting with a (hyper-)cubic lattice with 
$N=L^d$ nodes and helical boundary conditions, and keeping only a fraction $q$ of the $dN$
links.

On these coupled networks (each obtained by bond percolation with parameter $q$), we study a 
site percolation problem by retaining only a fraction $p$ of all nodes, calling the set of 
retained nodes ${\cal N}_p$. We define $\cal AB$-clusters as subsets of nodes $\in{\cal N}_p$ 
that are connected both in $\cal A$ and in $\cal B$. More precisely, assume 
that $C=\{i_1,i_2,\ldots i_m\}$ is a subset of nodes in ${\cal N}_p$. We call it a (connected $\cal AB$-)
cluster, if any two points $i \in C$ and $j\in C$ are connected by (at least) two paths:
one path using only links $\in \cal A$, and nodes only $\in C$, and another path using 
only links $\in \cal B$, also using nodes only $\in C$. Notice that we do {\it not} allow paths
that involve nodes outside $C$, i.e. $\cal AB$-clusters are `self-sustaining'. The ``order 
parameter" $S=m_{\rm max}/N$ is then the relative size of the largest $\cal AB$-cluster, for 
given $p$ and $q$.

To find these maximal clusters, we start with an empty initial configuration with no nodes but 
with a list of all possible links in $\cal A$ and$\cal B$,
and set $m_{\rm max}=0$. Then we add nodes one by one. Each time a new node $i$ is added,

(a) We check whether it is linked to any of the existing nodes. If it is not linked to any 
other node either by $\cal A$ or by $\cal B$ links, we simply insert the next node.

(b) Otherwise, we update the cluster structures in $\cal A$ and $\cal B$ separately by means
of the Newman-Ziff algorithm, and denote the sets of nodes linked to $i$ by $C_{\cal A}$ 
and $C_{\cal B}$. If one of them has size $\leq m_{\rm max}$, then $m_{\rm max}$ cannot 
increase and we insert the next node.

If not, we check whether the biggest $\cal AB$-cluster in $C_{\cal A}\bigcap C_{\cal B}$ 
can have a size $>m_{\rm max}$,
by following a cascade similar to that in \cite{Buldy}. If the cascade stops at a cluster 
size $>m_{\rm max}$, then $m_{\rm max}$ is increased. If it continues to a size 
$\leq m_{\rm max}$, the cascade is stopped and $m_{\rm max}$ is left unchanged. In either 
case, we then insert the next node.

(c) This process continues until a preset value $p_{\rm max}$ is reached. Stopping at $p<1$ 
is crucial for efficiency, as the algorithm slows down dramatically at large $p$.
We typically follow the evolution up to $p$ slightly above $p_c$ for 
all realizations, and follow it up to larger values of $p$ for successively fewer 
runs. This reflects the fact that simulations are slow for $p\gg p_c$, but fluctuations 
are also smaller, so that fewer samples are sufficient. 

For ER graphs the model can be simplified, since bond and site dilution 
both lead again to ER graphs. Hence we do not have to distinguish between them and can 
skip the site percolation part.
The order parameter $S = m_{\rm max}/N$ is then, in the limit $N\to\infty$, a unique 
function of the average degree $\langle k\rangle$. This function is easily found by 
arguments analogous to those for single networks. 

Consider an isolated ER network with average degree $\langle k\rangle = z$ in the regime where an
infinite cluster exists, i.e. where an infection has a non-zero chance to lead to an 
infinite epidemic. Let $S_i$ be the probability that node $i$ gets infected during this 
epidemic.  The probability that $i$ does {\it not} get infected is then 
\be
   1-S_i = \prod_{<ij>}(1-S'_j),     \label{Si}
\ee
where the product runs over all neighbors of $i$. Here, $S'_j$ is the probability that 
$j$ is infected, conditioned on it being picked as a node at the end of a link, and 
we used the fact that the graph is locally tree-like, so all $S'_j$ are independent.
For ER graphs the degree distribution is Poisson, and $S'$ and $S$ obey the same statistics. 
Averaging Eq.~(\ref{Si}) 
over all nodes and topologies gives then \cite{bollobas,newman}
\be
   1-S = \sum_k {e^{-z} z^k\over k!} (1-S)^k = e^{-zS},
\ee
where we dropped the index on $S$. Otherwise said, the probability $S$ that any site
is linked to the infinite cluster is $1-\exp(-zS)$. For two interdependent ER networks 
with average degrees $z_{\cal A}$ and $z_{\cal B}$, the chance to belong to the infinite 
${\cal AB}$-cluster is equal to the probability to be linked to it both via $\cal A$
and via $\cal B$, giving
\be
   S = (1-e^{-z_{\cal A}S})(1-e^{-z_{\cal B}S}).      \label{S2}
\ee
Although this is much simpler than the theory presented in \cite{Buldy}, it is exactly 
equivalent. It is generalized trivially to $>2$ interdependent networks \cite{Gao}, and
to other types of interdependencies \cite{Son}. 
If $z_{\cal A}=z_{\cal B}=z$, one finds only the solution $S=0$ for 
$z<z_c = 2.455407\ldots$, while a second stable solution $S>0$ exists for $z>z_c$. 
Just above threshold, $S_c = 0.511699\ldots$.

\begin{figure}
\begin{center}
\psfig{file=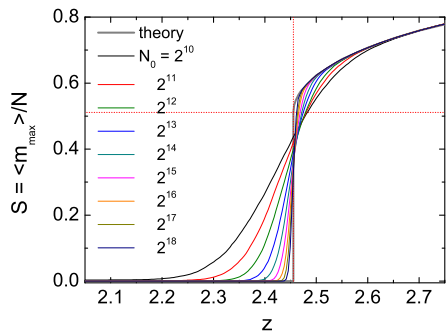,width=5.8cm, angle=0}
\caption{(Color online) Plot for $S = \langle m_{\rm max}\rangle /N$ against $z$, for two interdependent 
ER networks with degrees $z_{\cal A}=z_{\cal B}=z$. For technical reasons, each curve does not correspond
to a fixed value of $N$, but of $N_0=4N/z$. The grey curve is the solution of Eq.~(\ref{S2}). 
The intersection of the horizontal and vertical lines indicate 
the point $(z_c,S_c)$.} \label{Sz.fig}
\end{center}
\vglue -.4cm
\end{figure}

\begin{figure}
\begin{center}
\psfig{file=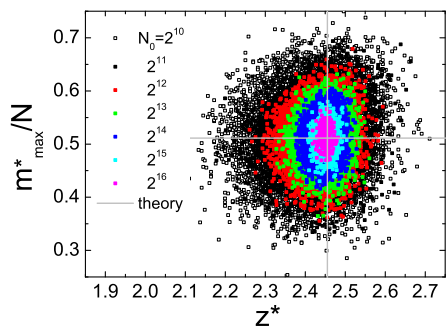,width=5.8cm, angle=0}
\caption{(Color online) Scatter plot for $m_{\rm max}/N$ against $z$, just after the largest jump 
in $m_{\rm max}$. Color corresponds to a fixed value
of $N$. The lines indicate the analytic prediction for the point $(z_c,S_c)$, according 
to Eq.~(\ref{S2}).} \label{scatter.fig}
\end{center}
\vglue -.4cm
\end{figure}

Results from our numerical simulations for ER graphs, using the algorithm outlined above, are shown in 
Figs.~\ref{Sz.fig} and \ref{scatter.fig}. Figure~\ref{Sz.fig} shows $S$ versus $z$ for networks of 
different sizes. Each curve is based on $10^4$ runs, except for the largest $N$.
The data indeed approach the theoretical curve (indicated in grey), as $N\to\infty$. 
While Fig.~\ref{Sz.fig} demonstrates that the theory gives the correct $z_c$, it is much harder 
to argue that it gives also the correct $S_c$. To see this, we notice that $m_{\rm max}/N$ makes in each 
run exactly one big jump, from $\approx 0$ to $\approx S_c$. The values of $z$ and
$S$ just after the jump are shown as scatter plots in Fig.~\ref{scatter.fig}. We see clouds of points
that are indeed centered near $z_c$ and $S_c$, and whose sizes decrease with $N$.

\begin{figure}
\begin{center}
\psfig{file=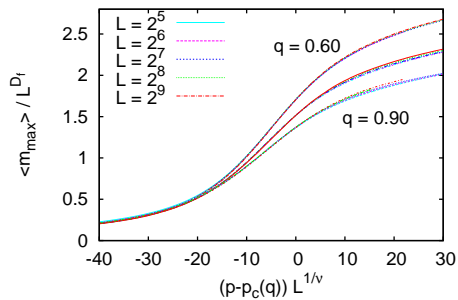,width=4.4cm, angle=270}
\caption{(Color online) Data collapse for $\langle m_{\rm max}\rangle/L^{D_f}$ against $(p-p_c)L^{1/\nu}$,
for 2-$d$ lattices.  Each set of curves corresponds to one value of $q$, while each curve within each set 
corresponds to a system size $L$. For this plot, $\nu=1.19$ and $D_f = 1.85$ were used, and the values of $p_c$ 
are $0.96025,0.77556$, and 
$0.6544$ for $q=0.6, 0.75$, and $0.9$. Due to universality of the scaling 
function $f(z)$ we can also collapse the three set of curves, by multiplying 
$m_{\rm max}/L^{D_f}$ and $(p-p_c)L^{1/\nu}$ by suitable $q$-dependent factors (data not shown).} \label{collapse_2d.fig}
\end{center}
\vglue -.4cm
\end{figure}

\begin{figure}
\begin{center}
\psfig{file=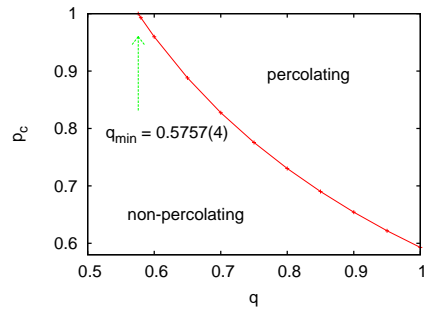,width=4.2cm, angle=270}
\caption{(Color online) Critical values $p_c$ versus $q$ for two coupled 2-$d$ lattices. Error bars 
are much smaller than the symbol sizes. Notice that the bond percolation threshold on the square 
lattice is $q_c=1/2$, thus the curve cannot extend below $q<0.5$. The transition is in the same 
universality class even for $q=q_{\rm min}=0.5757(4)$, where $p_c=1$ and the model simplifies, as no site
percolation is involved. For $q\to 1$, the model crosses over to OP.} \label{pc_2d.fig}
\end{center}
\vglue -.4cm
\end{figure}

For bond percolation on the square lattice, the OP threshold is at $q_c=1/2$ \cite{Stauffer}.
We therefore look for $\cal AB$-percolation in the parameter range $1/2 < q < 1$. 
We assume the usual finite size scaling (FSS) ansatz \cite{Stauffer}
\be
   \langle m_{\rm max} \rangle = L^{D_f} f[(p-p_c)L^{1/\nu}]\;,
\ee
where $\nu$ is the correlation length exponent, $D_f = d-\beta/\nu$ is the fractal dimension
of the incipient infinite cluster, and $f(z)$ is a smooth (indeed analytic) function. 
According to this {\it ansatz}, we expect a data collapse if we plot $\langle m_{\rm max} \rangle /L^{D_f}$
against $(p-p_c)L^{1/\nu}$. Three such data collapses are shown in Fig.~\ref{collapse_2d.fig}, 
each for a different value of $q$. 
Each of the three ``curves" in this figure are indeed several collapsed curves 
corresponding to different values of $L$ in the range $2^5$ to $2^9$, obtained from more than 
$10^6$ realizations for the smallest lattice and $\approx 10^4$ for the largest. For all
curves the same values of $D_f$ and $\nu$ were used, while $p_c$ depends of course on $q$.
The values of $p_c$ are plotted against $q$ in Fig.~\ref{pc_2d.fig}.

\begin{figure}
\begin{center}
\psfig{file=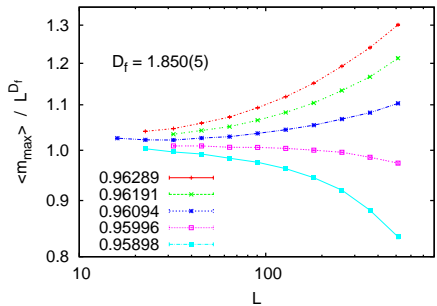,width=4.3cm, angle=270}
\caption{(Color online) Log-log plot of $\langle m_{\rm max}\rangle /L^{D_f}$ against $L$, for 2-d
interdependent percolation at $q=0.60$ and at fixed values of $p$. At the critical point 
($p_c = 0.96025(20)$) we expect a 
straight line. The value of the fractal dimension $D_f$ is chosen such that this line 
is horizontal.}     \label{df_2d.fig}
\end{center}
\vglue -.4cm
\end{figure}

\begin{figure}
\begin{center}
\psfig{file=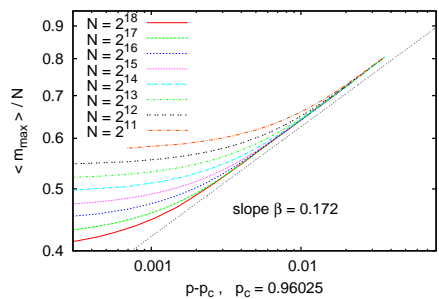,width=4.3cm, angle=270}
\caption{(Color online) Log-log plot of $S=\langle m_{\rm max}\rangle/L^2$ against $p-p_c$, for 2-d 
interdependent percolation with $q=0.60$. The slight upward curvature for large $p-p_c$ indicates the 
limit of the critical region, while the upward curvature for $p- p_c\to 0$ is due to finite
size corrections. }   \label{beta_2d.fig}
\end{center}
\vglue -.4cm
\end{figure}

The fact that data collapse was obtained in Fig.~\ref{collapse_2d.fig} for $q$-independent 
values of the exponents indicates that these exponents are universal for $q_{\rm min} \leq q < 1$. But a closer 
inspection of Fig.~\ref{collapse_2d.fig} shows that the quality of the collapse deteriorates
as $q\to 1$, due to the expected cross-over to OP (for $q\to 1$, $\cal A$ and $\cal B$
become identical, and the problem crosses over to OP). Thus we use data for $q=0.6$ for 
more detailed analyses. 
Figure~\ref{df_2d.fig} shows that $m_{\rm max} \sim L^{D_f}$ for $p_c = 0.96025(20)$, 
with $D_f=1.850(5)$, while Fig.~\ref{beta_2d.fig} shows that $m_{\rm max} \sim (p-p_c)^\beta$
in the limit $L\to\infty$, with $\beta = 0.172(2)$ (for a plot with higher resolution see 
the supplementary material (SM)). Both exponents are clearly different 
from the values for OP. Indeed, $\beta$ is larger than the value $5/36=0.1389$ for OP,
showing that the transition is not more abrupt than in OP, as claimed in \cite{Buldy}, but
less so!

For $d=3$ we also studied systems of up to $2^{18}$ sites, with roughly the same number of 
realizations as for 2d, and with similar results (see the SM for 
details): There are also important corrections to scaling, if $q$ is taken too large, 
but they decrease strongly when $q$ is taken as small
as possible. For $q=0.40$ we obtain $p_c = 0.871(1), \beta=0.51(1), \nu = 0.86(1),$ and 
$D_f = 2.40(1)$. These values satisfy (like the 2-d exponents) the scaling relation $D_f=d-\beta/\nu$, 
and again they are 
incompatible with OP (where $\beta=0.4170(3), \nu = 0.8734(5), D_f = 2.5226(1)$ \cite{Deng}). 
As in 2-d, $\beta$ is clearly larger than in OP, indicating that the 
transition is again less sharp, rather than more abrupt.

In summary, we have shown that coupling two interdependent networks does not generically 
make the percolation transition more abrupt or discontinuous. Rather, the outcome 
depends on the network topologies. Real networks (e.g. transportation,
telephone, ...) often are locally embedded in space, thus their behavior might 
resemble more that of regular lattices than that of small world networks. The reason 
why the claim of \cite{Buldy} does not hold universally is not that the cascade picture
breaks down for local networks. Rather, cascades are an essential ingredient in any 
spreading phenomena on any network, and it depends on the topology whether or not their 
effects are enhanced by the coupling between different networks.

In the present paper we have only studied two statistically identical networks. It 
is an open question what happens, say, when a diluted 2-d lattice is fully coupled 
to an ER network or a scale-free one. Also, one might think of more than 2 interdependent 
networks \cite{Gao}. In view of possible applications, one should also study networks that
are semi-locally embedded in 2-d space. The latter could also be used to study the 
cross-over from networks with local connections (as in 2-d lattices) to global
(e.g. ER) networks. A priori, one might expect that there exists a tricritical point
between these two extremes, or that one of them is unstable against even infinitesimal
perturbations.

\end{document}


\title{Supplementary Material for ``Are percolation transitions always sharpened by making networks
interdependent?"}
\author{Seung-Woo Son} \affiliation{Complexity Science Group, University of Calgary, Calgary T2N 1N4, Canada}
\author{Maya Paczuski} \affiliation{Complexity Science Group, University of Calgary, Calgary T2N 1N4, Canada}
\author{Peter Grassberger} \affiliation{Complexity Science Group, University of Calgary, Calgary T2N 1N4, Canada}
\date{\today}

\maketitle

In this supplementary material, we present several plots. They illustrate claims for which the data are not 
shown in the main paper, or they are plotted in different ways. All necessary
information to explain the plots is given in the figure captions.

\begin{figure}[htbp]
\psfig{file=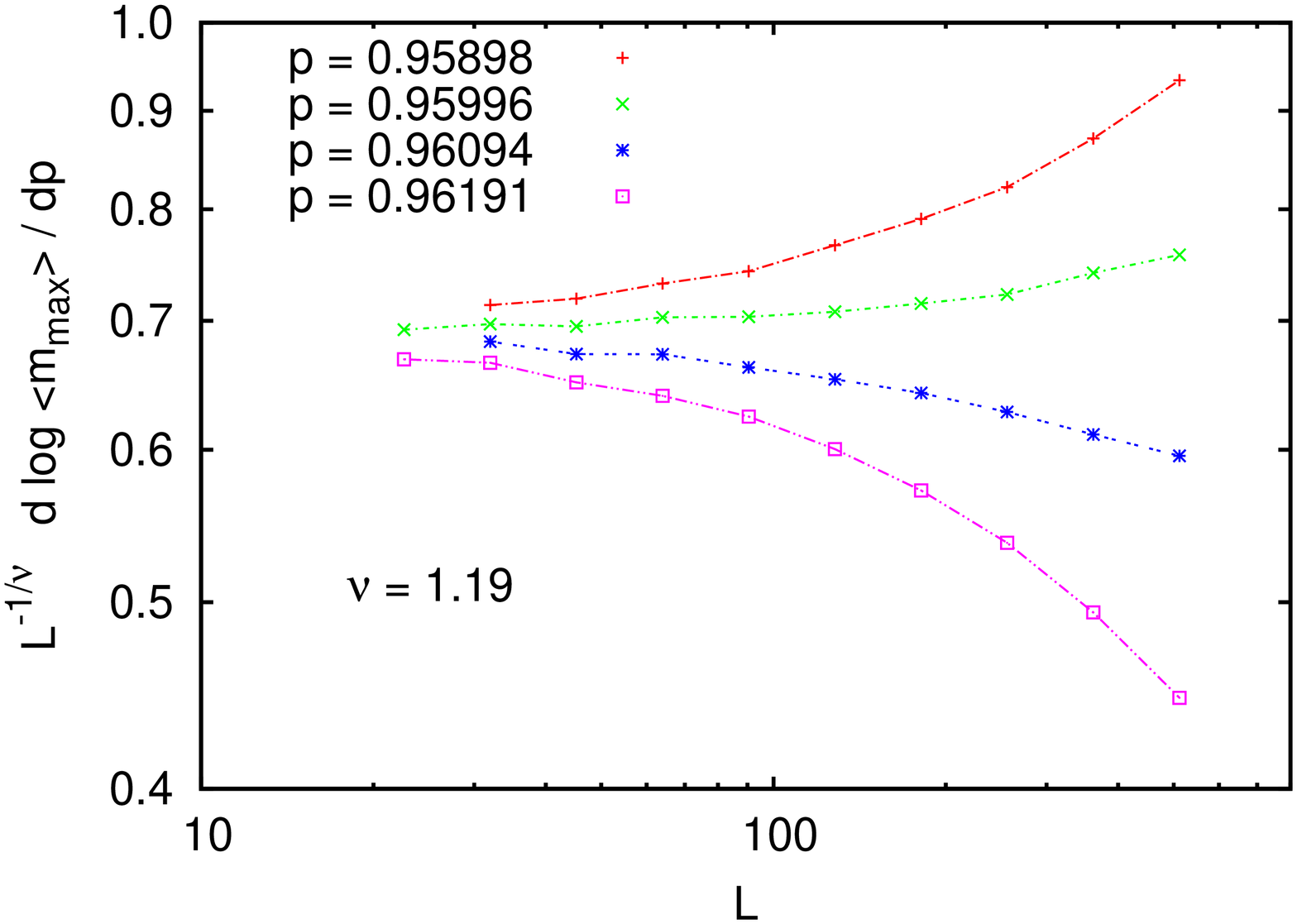,width=7.9cm, angle=270}
\caption{(Color online) Log-log plot of $L^{-1/\nu} d \log \langle m_{\rm max}\rangle / dp$, where the derivative with 
   respect to $p$ is approximated by a finite difference quotient with $dp=1/1024$. According to the scaling ansatz 
   Eq. (4), this should be a constant at $p=p_c$. It is indeed constant (independent of $L$) at $p=0.96025$ within the 
   expected accuracy (we do not show error bars, since the derivative is obtained from correlated measurements, 
   making any error estimation difficult), if we assume $\nu=1.19$. Within the quoted error bars on $D_f$ and $\beta$,
   this agrees with the hyperscaling relation $D_f = 2-\beta/\nu$. Notice that the OP value $\nu=4/3$ is definitly
   excluded.} 
   \label{fig1}
\end{figure}

\begin{figure}[htbp]
\psfig{file=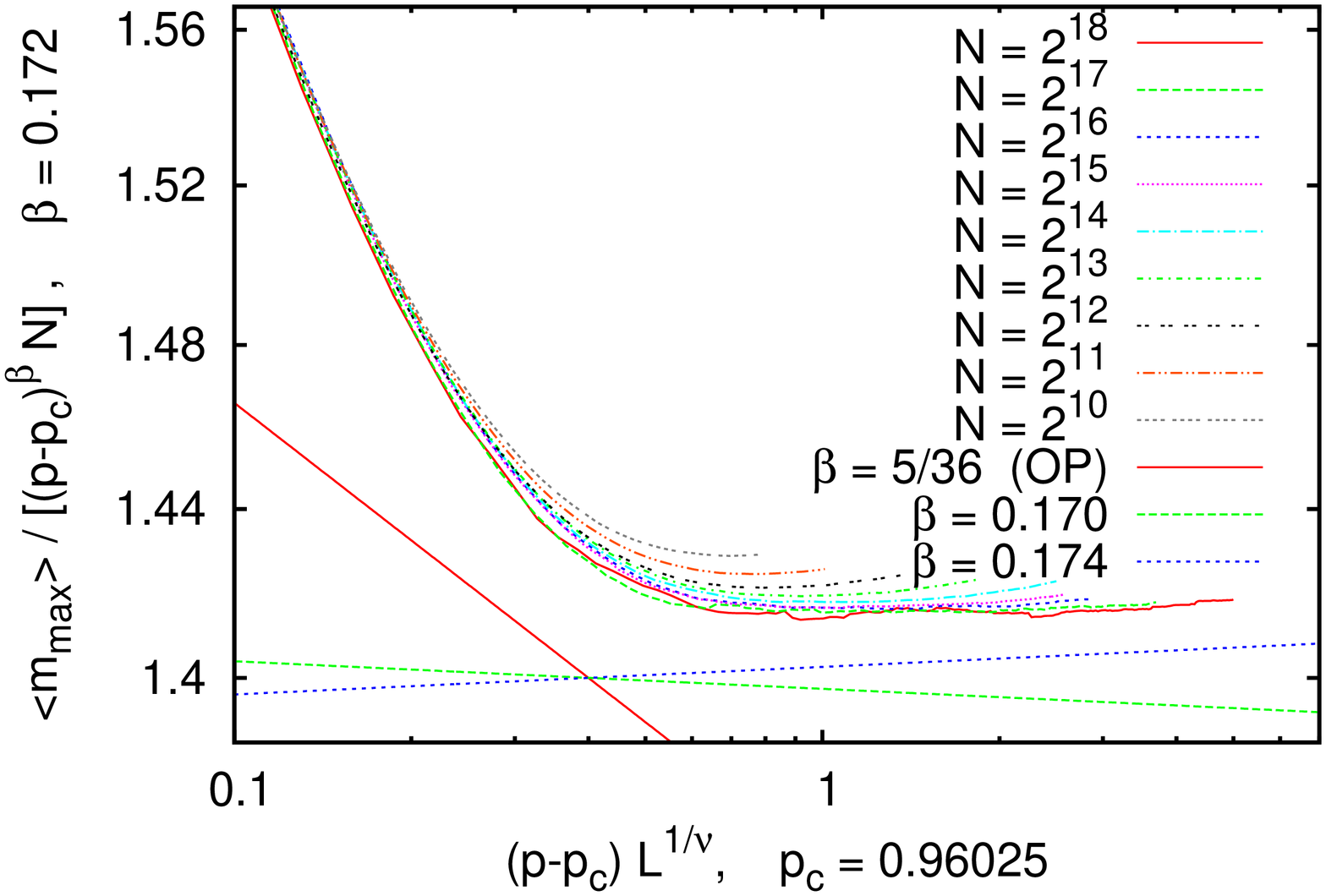,width=7.9cm, angle=270}
\caption{(Color online) Log-log plot of $\langle m_{\rm max}\rangle /[(p-p_c)^\beta N]$ versus $(p-p_c)L^{1/\nu}$, 
   with $p_c$ and $\beta=0.172$ as given in the paper and $\nu=1.19$ as obtained from the hyperscaling relation, with 
   $D_f$ as given in the paper. If these values and the scaling ansatz Eq. (4) were exact, the curves would perfectly 
   collapse onto a single curve that is horizontal in the region $O(1) < (p-p_c)L^{1/\nu} < O(L^{1/\nu})$. Deviations
   mainly result from finite size corrections. The three straight lines indicate the slopes we should observe in the 
   power law region,
   if $\beta$ were different from the nominal value $\beta=0.172$. The full (red) line corresponds to $\beta = 5/36$
   in OP, and is solidly excluded. The two dashed lines correspond to $\pm 1\sigma$. Notice the very small range
   of the plotted data in the $y$ direction, leading to enormous blow-up of errors and to a much higher significance 
   than in Fig.~6 of the main text.}
   \label{fig2}
\end{figure}

\begin{figure}[htbp]
\psfig{file=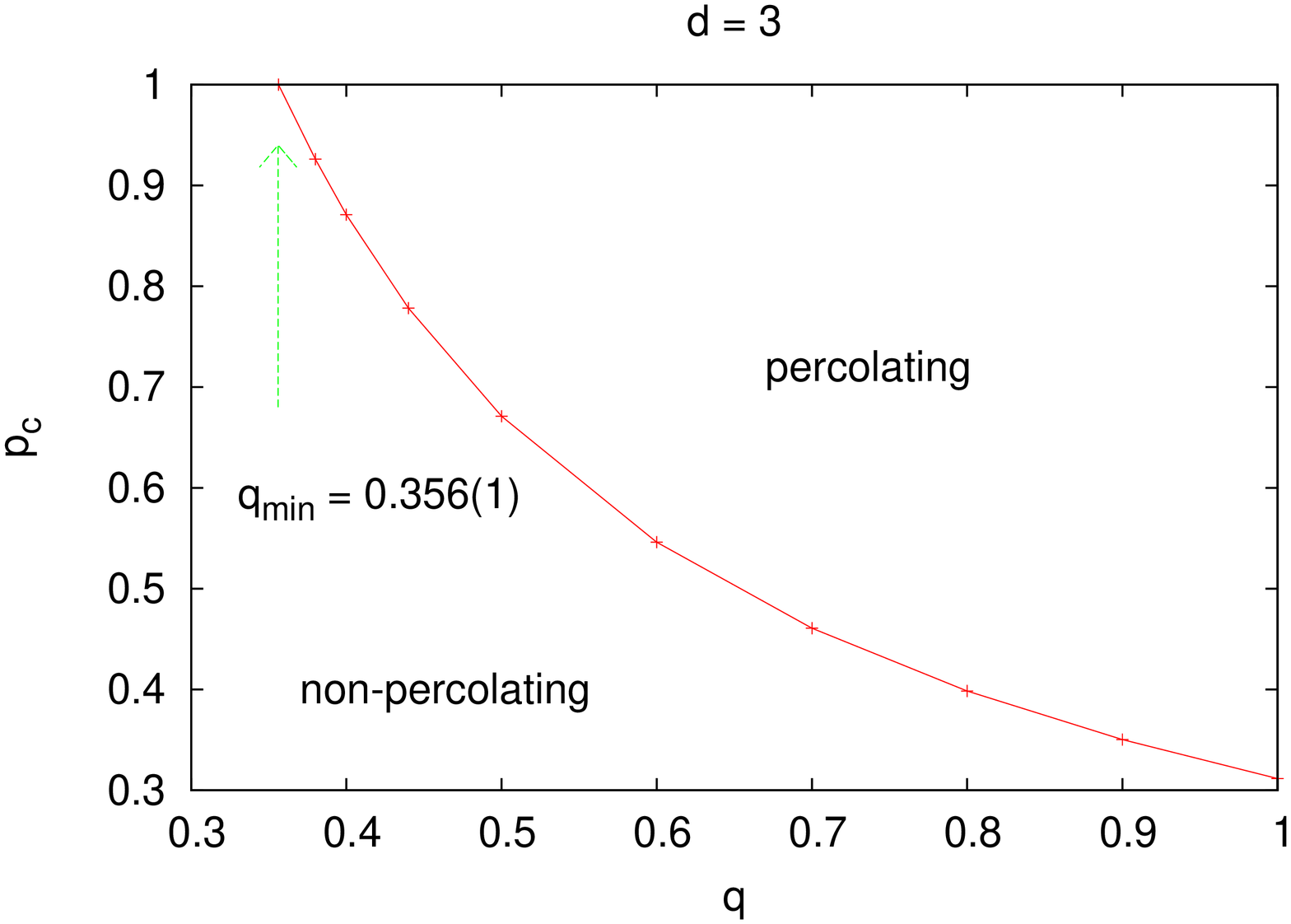,width=7.9cm, angle=270}
\caption{(Color online) Critical values $p_c$ versus $q$ for coupled 3-$d$ lattices. This plot is to be
   compared to Fig.~4 of the main paper, where results are shown for two dimensions. This time a lower bound 
   on $q_{\rm min}$ is given by the value of $p_c=0.248812\ldots$ of 3-$d$ bond percolation, while $p_c$ for $q=1$
   equals the critical value $0.31160\ldots$ of 3-$d$ site percolation. As in two dimensions, we suggest 
   that the transition is in the same universality class in the entire range $q_{\rm min} \leq q < 1$, while
   it is in the OP universality class for $q=1$.}
   \label{fig3}
\end{figure} 

\begin{figure}[htbp]
\psfig{file=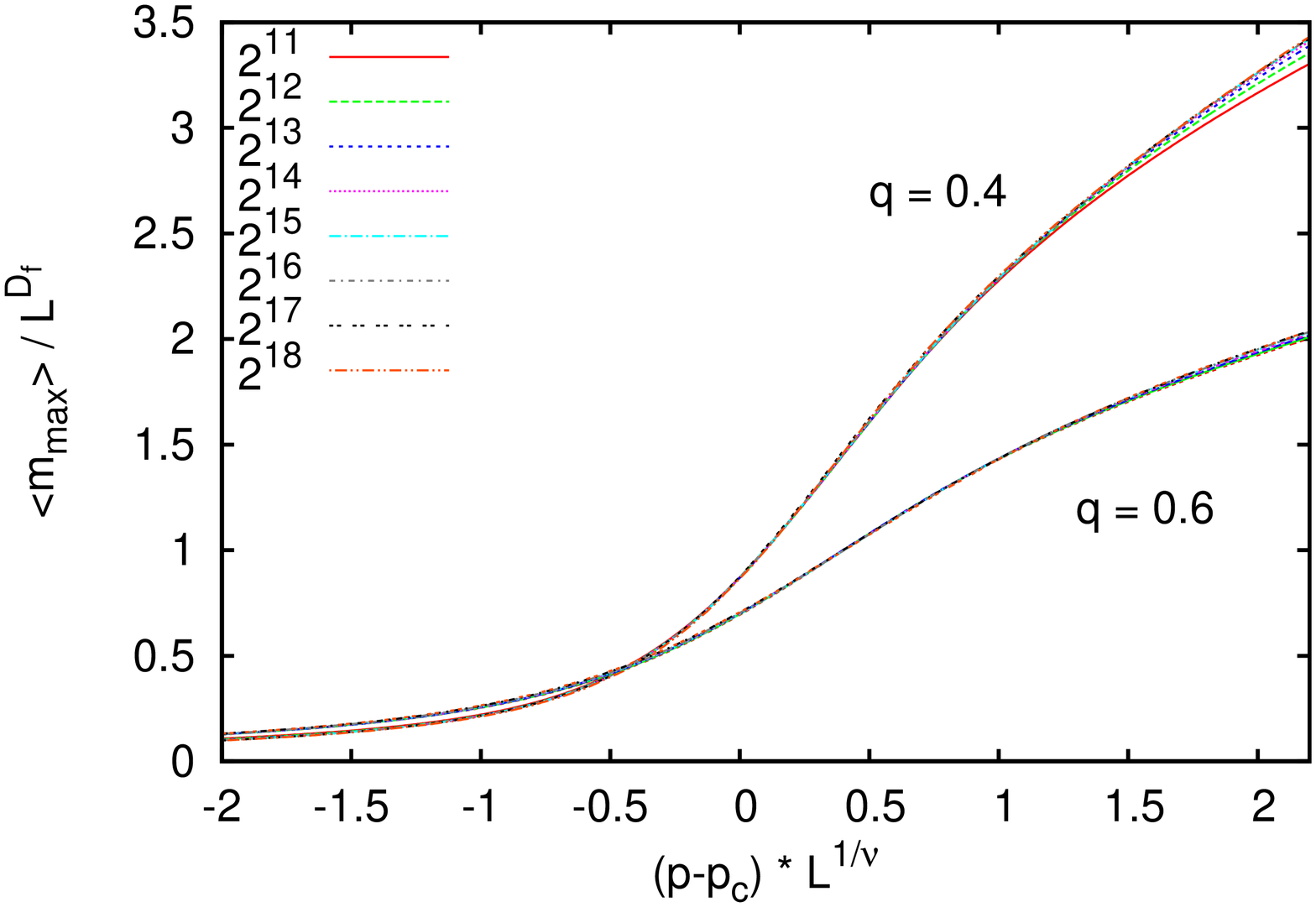,width=7.9cm, angle=270}
\caption{(Color online) Data collapse for $\langle m_{\rm max}\rangle /L^{D_f}$ versus $(p-p_c)L^{1/\nu}$ 
   for 3-$d$ lattices. As in Fig.~3 of the main paper, each set of curves corresponds to one value of $q$.
   In contrast to the 2-$d$ case, we needed however slightly different critical exponents in order to 
   obtain good collapses for both values of $q$. More precisely, we used for $q=0.4$ the exponent values
   $\beta=0.51$ and $\nu=0.86$ 
   quoted in the main paper, while we used $\beta =0.446 $ and $\nu = 0.77$ for $q=0.6$. The values of 
   $p_c$ used in this plot are 0.871 (for $q=0.4$) and 0.5464 (for $q=0.6$). Notice that both sets of curves in the present 
   figure show visible deviations from a perfect collapse, but -- in contrast to what might be suggested by
   the figure -- we claim that the effective exponents obtained for $q=0.4$ are much closer to the true
   ones. This conclusion is mainly based on the following figures (Figs. S6 to S9) which would show 
   substantially more scaling violations for $q=0.6$ than they do for $q=0.4$. We interpret this as a 
   manifestation of the cross-over from OP, that should be much more important at $q=0.6$ than at $q=0.4$.}
   \label{fig4}
\end{figure}

\begin{figure}[htbp]
\psfig{file=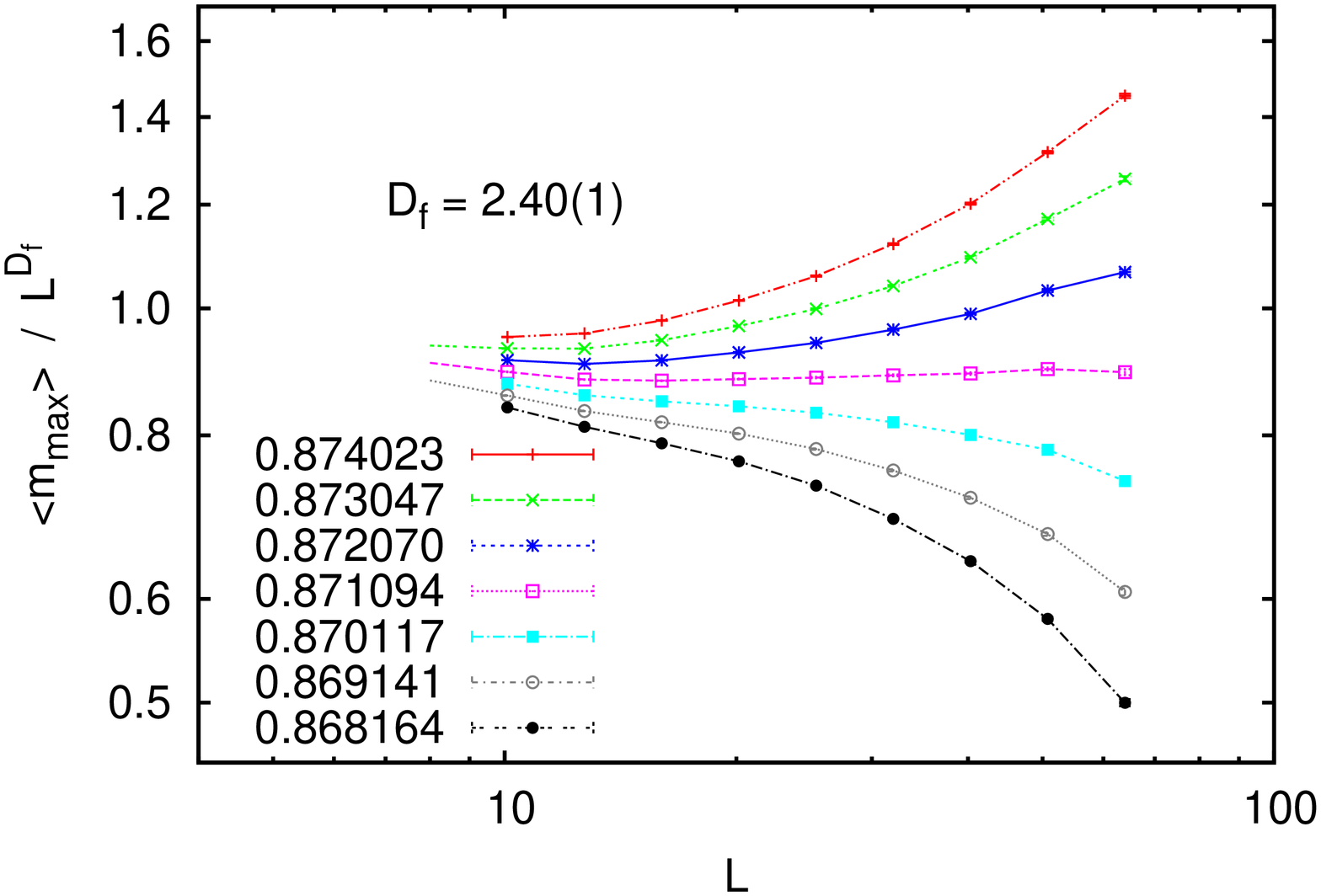,width=7.9cm, angle=270}
\caption{(Color online) Log-log plot of $\langle m_{\rm max}\rangle /L^{D_f}$ versus $L$ for 3-$d$ 
   interdependent percolation at $q=0.40$ and fixed values of $p$ (analogous plot to Fig.~5 of the main
   paper). The fractal dimension is chosen such that the curve for $p=p_c$ is horizontal, with 
   $p_c = 0.871$.}
   \label{fig5}
\end{figure}

\begin{figure}[htbp]
\psfig{file=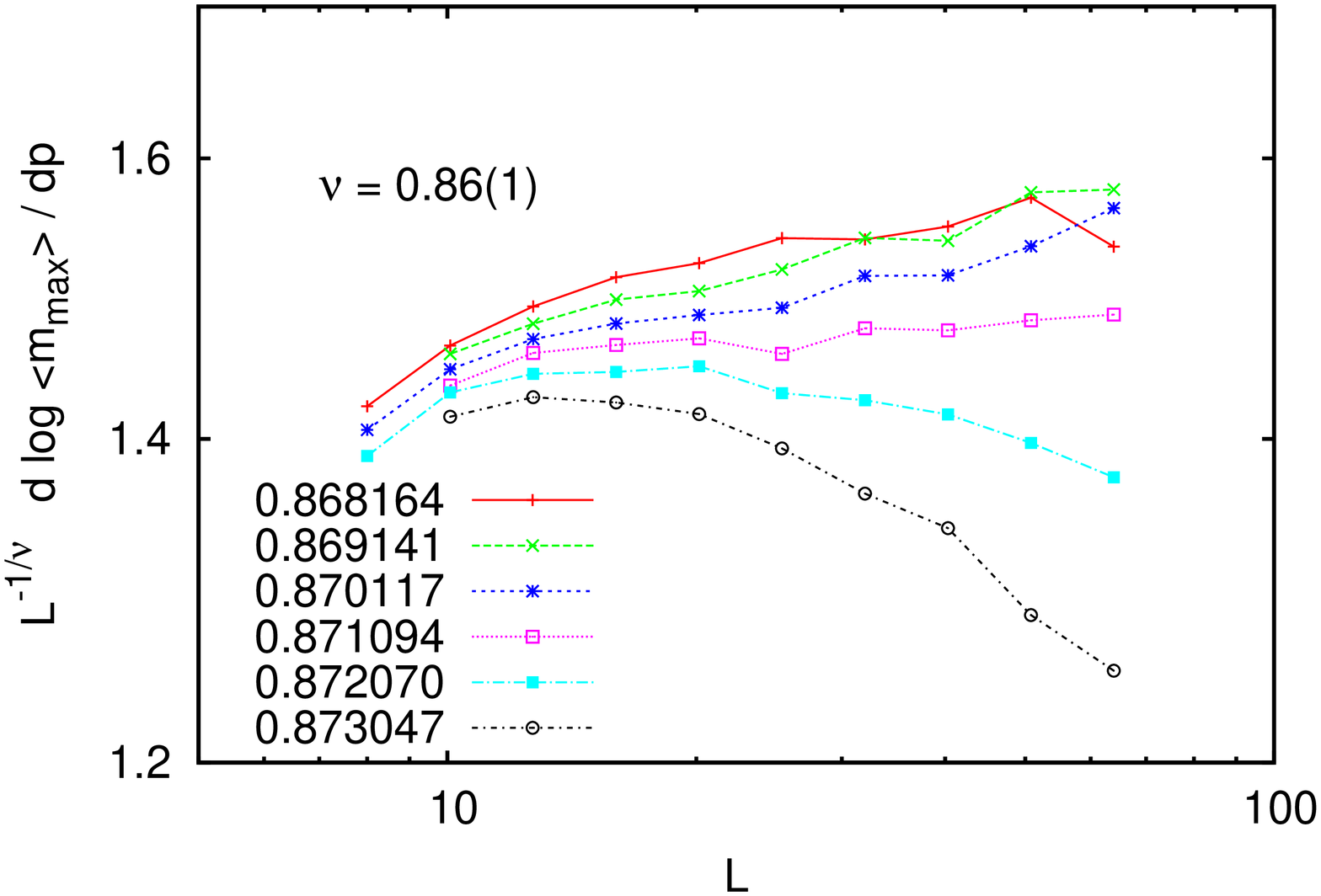,width=7.9cm, angle=270}
\caption{(Color online) Analogous plot to Fig.~S1, but for $d=3$ and $q=0.4$. We see strong finite 
   size corrections for small values of $L$. For $q=0.6$, scaling violations would 
   extend to the largest values of $L$ (see Fig. S7).}
   \label{fig6}
\end{figure}

\begin{figure}[htbp]
\psfig{file=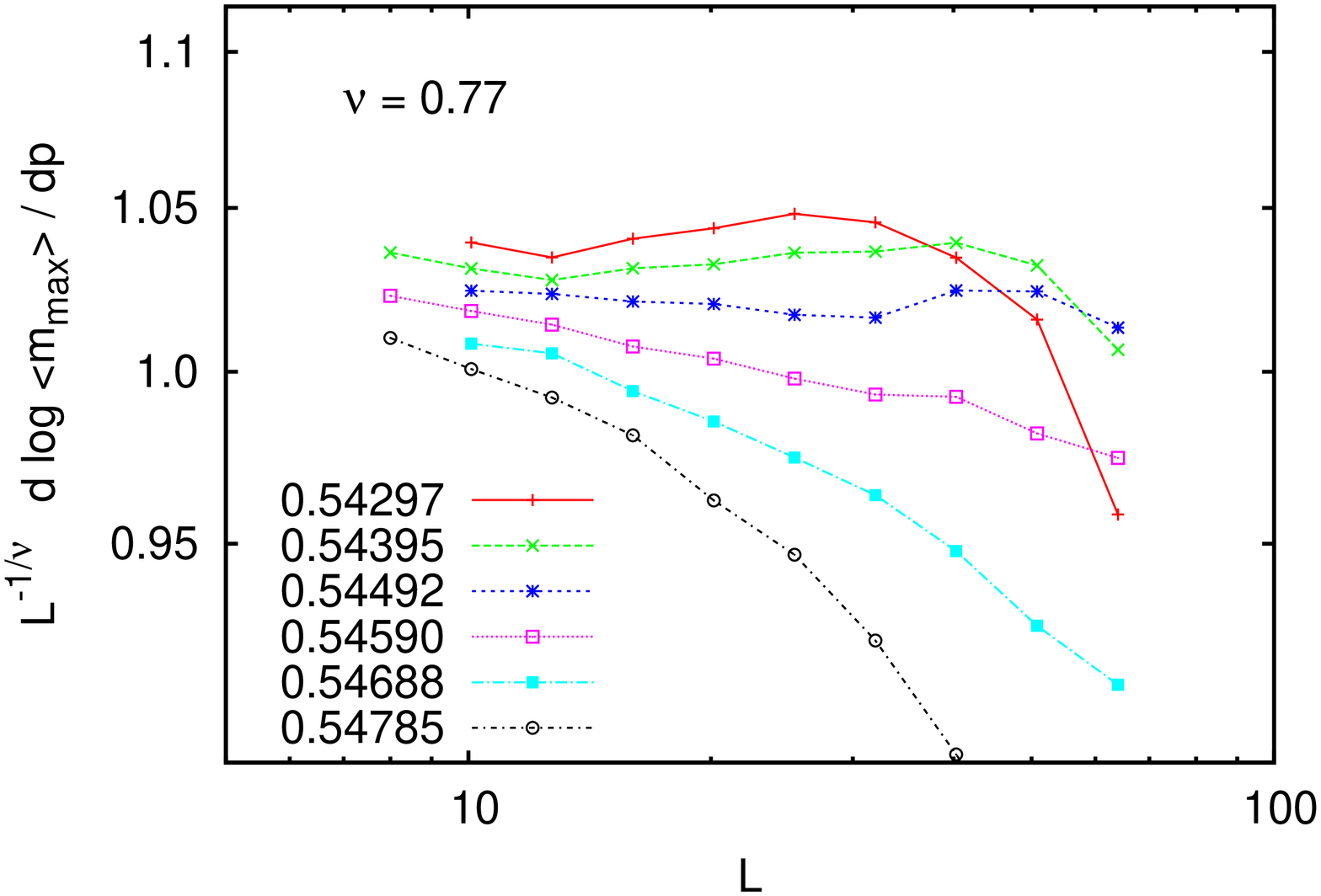,width=7.9cm, angle=270}
\caption{(Color online) Same as Fig.~S6, but for $q=0.6$. The value for $\nu$ is that 
   used in Fig. S4 to obtain the best data collapse. We see now that this good data 
   collapse was spurious. Either $\nu$ or $p_c$ as obtained in Fig. S4 are wrong, but 
   most dramatic is the unphysical curve crossings for $p<p_c$. While the consistency 
   of $\nu$ and $p_c$ could be improved at the cost of deteriorating somewhat the data 
   collapse in Fig. S4, the curve crossings for $p<p_c$ cannot be avoided.}
   \label{fig7}
\end{figure}

\begin{figure}[htbp]
\psfig{file=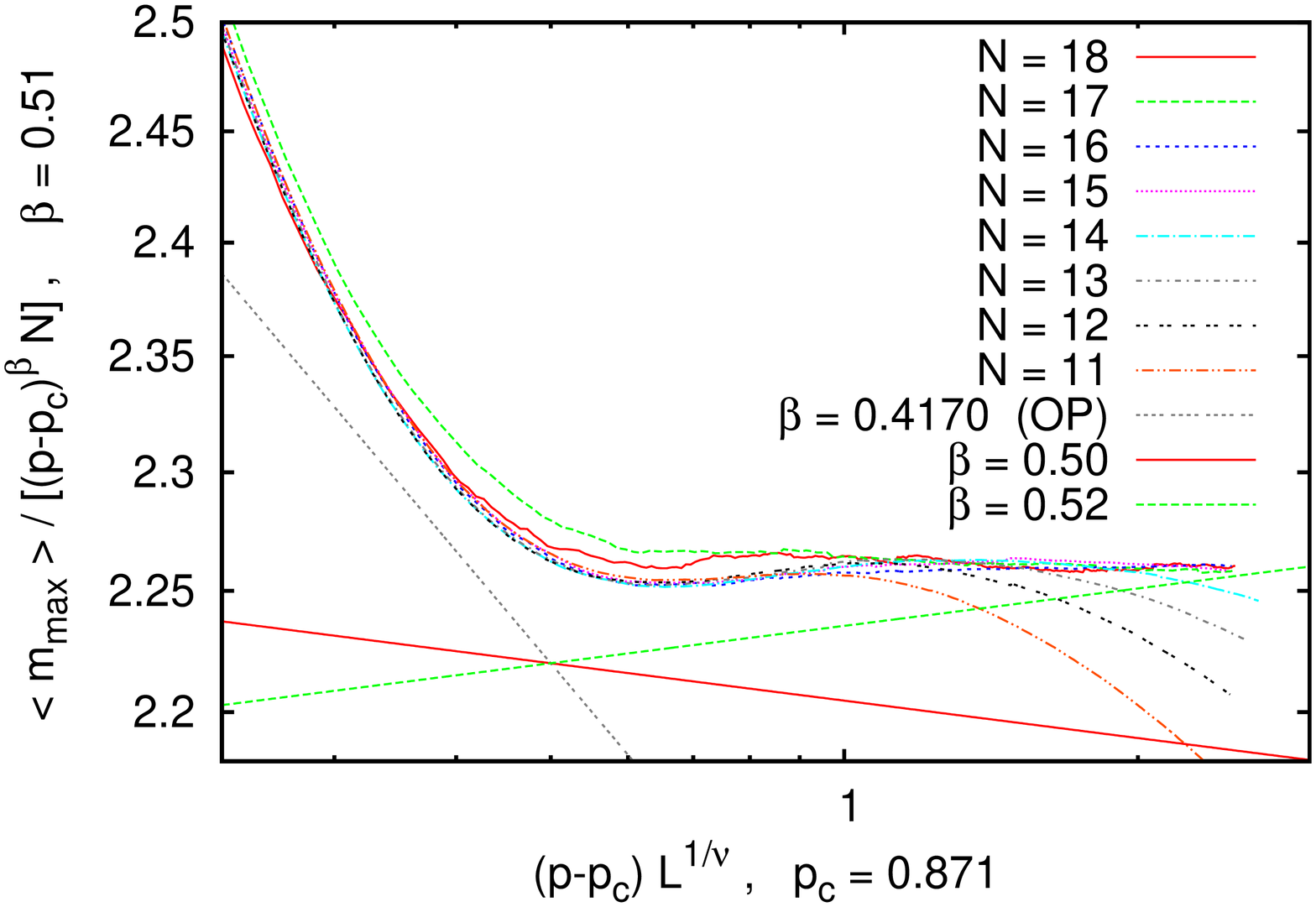,width=7.9cm, angle=270}
\caption{(Color online) Analogous plot to Fig.~S2, but for $d=3$ and $q=0.4$. The values
   of $p_c$ and of the critical exponents are those given in the caption to Fig. S4 and 
   in the main text.}
   \label{fig8}
\end{figure}

\begin{figure}[htbp]
\psfig{file=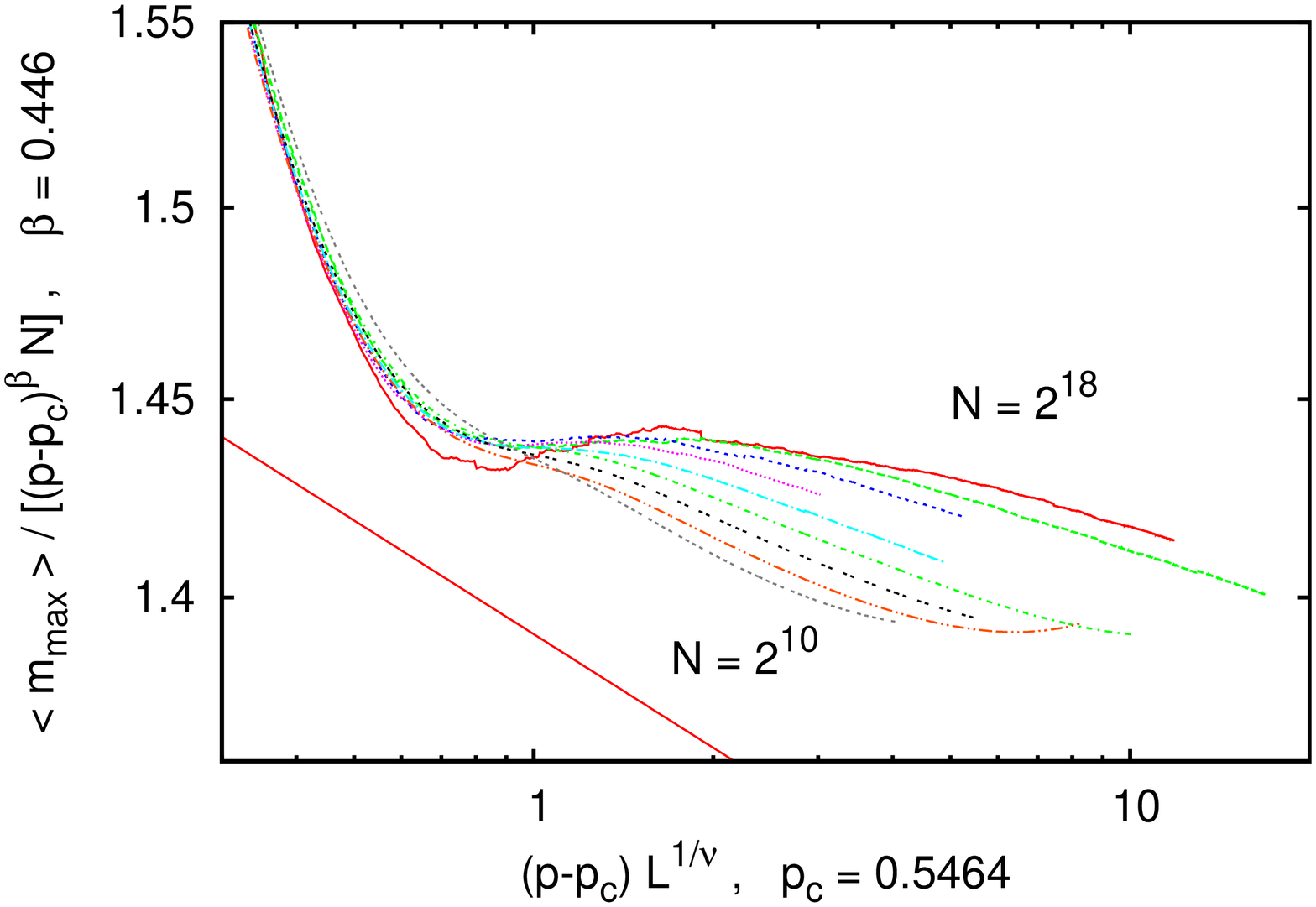,width=7.9cm, angle=270}
\caption{(Color online) Analogous plot to Fig.~S8, but for $q=0.6$. The values
   of $p_c$ and of the critical exponents are those given in the caption to Fig. S4  
   and used also in Fig. S7. The straight line represents the slope expected for OP 
   (since we do not assign error bars to the value of $\beta$ used in this plot, we 
   do not show the two other straight lines that were shown in Figs. S2 and S8). Again 
   we see that the data collapse in Fig. S4 for $q=0.6$ is spurious (it is restricted 
   to $(p-p_c) L^{1/\nu}<1$), and that corrections to scaling are huge at $q=0.6$. 
   Similar but less dramatic scaling violations were also seen in $d=2$, if $q=0.75$ 
   was used instead of $q=0.6$ (data not shown). This supports our conclusion that 
   scaling violations are due to cross-over from OP.} 
   \label{fig9}
\end{figure}